# Absence of Small Dust Cloud Particles Transiting the White Dwarf J0328-1219

Bruce L. Gary[1] and Thomas G. Kaye[2]

[1] Hereford Arizona Observatory, Hereford, AZ 85615

[2] Raemor Vista Observatory, Sierra Vista, AZ 85650

Abstract

The transiting dust clouds that orbit the white dwarf J0328-1219 are devoid of small particles (< 0.1 micron). Observations show that fade amount doesn't depend on wavelength. This finding resembles a similar observation for white dwarf WD 1145+017, but the explanations for an absence of small particles in the two white dwarf systems may differ due to their different distances from the star.

1. Introduction

ZTF J0328+1219 (hereafter J0328) is a white dwarf (WD) with orbiting dust clouds producing multiple fade events ("dips") each orbit. In many respects J0328 resembles WD 1145+017 (hereafter WD1145), which is the prototype for WDs with dust cloud dips. With eight candidates belonging to this WD category they still defy understanding: "what mechanism produces the dust clouds?"

For both WD1145 and J0328 the dust cloud orbits are so small that the WD's gravity dominates everything exterior to the planetesimal's surface (assuming its shape conforms to a Hill sphere). This "tidal disruption" means that escape velocity is zero over the entire planetesimal's surface, and this can lead to a gradual dismantling of the planetesimal if the orbit is slowly shrinking.

This tidal disruption model is attractive, but observed episodic activity, and phase drifting behavior, suggest that collisions occur between previously separated fragments of the planetesimal. Once a dust cloud is created it has a perplexing ability to persist for months with the same depth and length structure. This behavior implies the presence of steady-state production and loss processes and is contrary to single creation event followed by dissipation due to Keplerian shear along the orbit.

Both J0328 and WD1145 have been observed to have more than a dozen dust clouds in orbit. The two systems differ, however, in the deepest dip depth observed, only 18 % for J0328 and 65 % for WD1145. Overall dip activity versus date is even more dramatic: J0328 has maintained a constant activity level (fractional loss of flux per orbit period) for the past 4 years, whereas WD1145's activity level rose abruptly in 2015 to ~ 12 % and has decreased exponentially every year since to levels that are no longer measurable (< 0.2 %).

Orbit periods of 4.5 hours (for WD1145) and 10 hours (for J0328) means that performing a photometric light curve (LC) during a dust cloud orbit is expensive when using professional observatories. V-mags for WD1145 and J0328 are 17.2 and 16.6, so performing LC photometry is a faintness challenge for non-professional observers.

2. Observations

The Hereford Arizona Observatory (HAO) is operated by BLG and consists of an equatorially-mounted 0.4-m AstroTech Ritchey-Chretien telescope with an SBIG ST-10XME CCD camera. The Raemor Vista Observatory consists of an alt/az 1.1-m telescope with a back-illuminated Apogee CCD camera constructed and operated by TGK. Both observatories are located in Southern Arizona.

The HAO 0.4-m telescope employed a g' filter (464 nm) with exposure times of 60 seconds. MaxIm DL was used for image calibration, and an Excel program created by author BLG for exoplanet photometry was used for optimizing MaxIm DL photometry files. The RVO 1.1-m telescope employed an R filter (641 nm) with exposure times that were usually 60 seconds. AstroImageJ was used for image calibration and analysis.

3. Light Curve Dip Depth Measurements

Observations using the g'/R pair of filters were obtained on 13 dates between 2023 December 03 and 2024 January 17. Each LC was fitted using "asymmetric hyper-secant" (AHS) functions. Comparing the LC for each filter requires a subjective LC normalization, and this was done with an emphasis on the "out of transit" portions. The deepest dips were considered for inclusion in the following analysis, which consisted of determining the average depth for each filter's data that was between the half-power times for the dip's AHS function. The 32 depth ratios that were accepted for analysis are shown in the following figure..

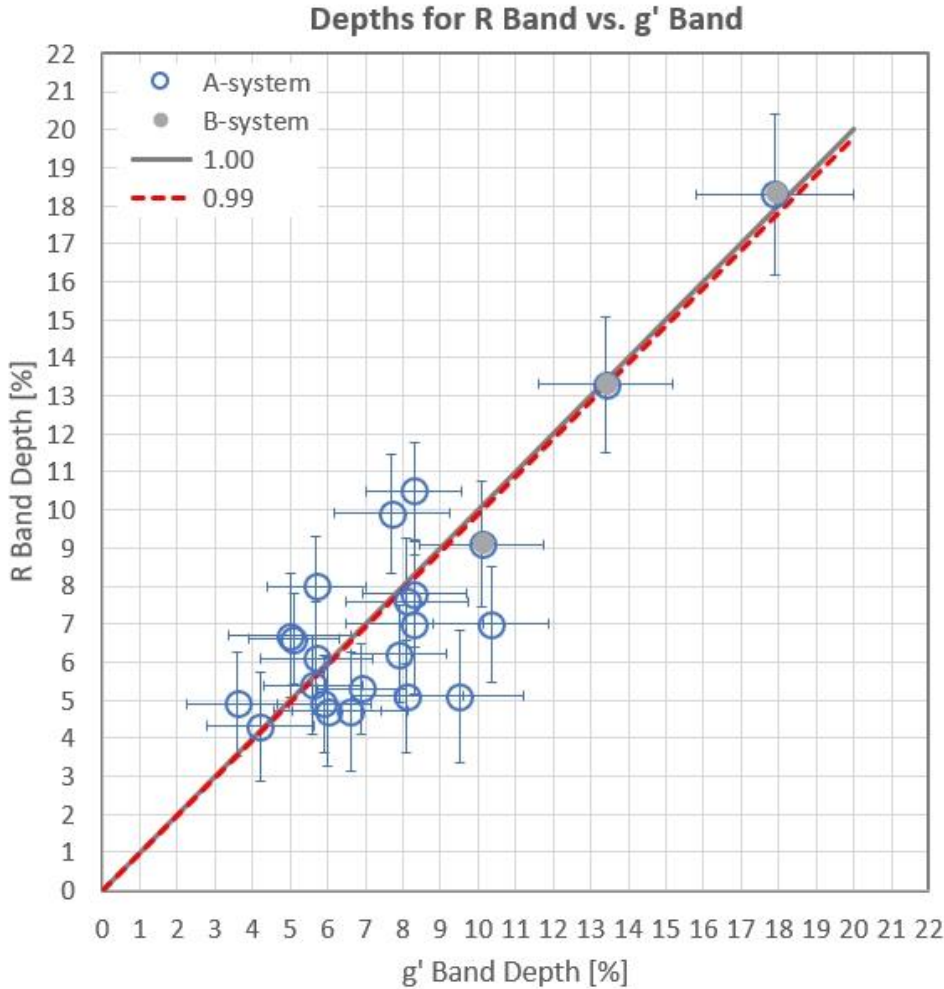

**Figure 1.** Dip depths at R band versus depth at g' band for 32 dip fades. All but 3 dips belong to the A-system orbit with P = 9.942 hours; 3 dips belong to the B-system orbit with P = 11.17 hours. The best-fit slope is 0.990 ± 0.055.

In this figure a distinction is made for dips that belong to the A- versus B-system orbits. Even though only three dips have a B-system periodicity they appear to exhibit the same lack of wavelength dependence that is strongly favored for A-system dips.

3. Discussion

Vanderbosch *et al*. (2021) established J0328's resemblance to WD1145 and stated that Zwicky Transient Factory r'-band measurements (which were the basis for identifying J0328) showed slightly deeper dips at r' band than g' band, which is counter to expectation for any plausible dust "particle size distribution" (PSD). The article added the following clarification: "*We note, however, that the ZTF g and r band measurements are not made simultaneously and cover a broad time baseline, which may impact the observed amplitude correlation. Simultaneous multi-band observations would be ideal for addressing this important question.*"

The measurements reported here answer the appeal for simultaneous two-wavelength measurements. There are two ways for a dust cloud to produce dips with the same depth at different wavelengths: 1) all particles are large (particle circumference > the longest wavelength), and 2) particles (of any size) are so numerous that their projected area blocks all light (for all line-of-sights to the star). Option 2 is difficult to fulfill when dips depths are << 100 % because any plausible dust cloud will have line-of-sight number densities that go to small values at cloud edges in a gradual manner. The J0328 dip depths range from 4 to 18 %, so option 2 can be disregarded. Mie scattering theory has been used to calculate extinction versus wavelength for plausible PSD assumptions (see Xu *et al.*, 2018). The measurements reported here have an average that differs from 1.00 by < 1-sigma, so we can state that J0328 dip depths vary with wavelength in accordance with Mie scattering.

An analysis of dip depths versus wavelength for WD1145 was published by Xu *et al.* (2018) showing that depth did not vary with wavelength from optical to 4.5 micron. The article speculated that the absence of small particles could be due to their low emissivity at their blackbody wavelengths. Since large dust particles orbiting WD1145 with a 4.5-hour period would be at temperatures slightly below where most minerals sublimate, the additional heating for small particles would increase their temperatures above most mineral's sublimation points. Hence, there should be no small particles in the WD1145 dust clouds, whether they ever existed before over-heating.

The situation for J0328 is different. Large particles in a larger 9.9-hour orbit should be heated to ~ 400 K. Any additional heating of small particles due to the low emissivity argument would be insufficient for reaching sublimation temperatures.

Is it reasonable to expect any dust clouds orbiting a star to have a PSD that includes particles smaller than ~ 0.1 micron? Yes, KIC 8462852 (aka "Tabby's Star") has two dust cloud components, and they exhibit r'/g' depth ratios significantly < 1 (Gary, private communication). The short dips (with a timescale of days) have r'/g' depth ratios of 0.60, which can be easily accounted for using plausible PSD models. The slowly varying "torus" component (month timescales, representing an "out-of-transit" level for analyzing the short timescale dips) exhibits an even more extreme depth ratio of ~ 0.45. This is close to the Rayleigh scattering limit for the two wavelengths involved (i.e., all particles in this "torus" component are smaller than 0.1 micron).

Light pressure cannot account for the presence of small particles in KIC 8462852 dust clouds and their absence for the two WD systems. This is because the ratio of light pressure force to gravity, commonly referred to as β, is proportional to a star's luminosity. For main sequence stars (earlier than spectral class K) β > 1 for dust with radii between 0.05 and 1 micron, and for relevant mineralogy. WDs have such low luminosities that their β values are <<1, so radiation forces cannot account for an absence of small particles.

Why, then, are small particles absent from the J0328 dust clouds? Could small particles simply not be created by whatever mechanism produces the dust clouds? The absence of small dust particles may eventually contribute to an understanding of the mechanism that produces WD dust clouds.

4 Acknowledgements

We thank Dr. Saul Rappaport for help in the interpretation of the statistics associated with Fig. 1.